\documentclass[aps,preprint,nofootinbib,floatfix,superscriptaddress]{revtex4}
\usepackage{color}
\usepackage{slashed}
\usepackage{amsmath,amssymb,graphicx,xcolor,mathtools,float}
\usepackage{graphicx}
\usepackage{epstopdf}
\usepackage[colorlinks=true,linktocpage=true,linkcolor=red,citecolor=blue]{hyperref}
\def\be {\begin{equation}}
\def\ee {\end{equation}}
\def\nn {\nonumber}
\def\bea {\begin{eqnarray}}
\def\eea {\end{eqnarray}}

\newcommand{\om}{\omega}  
\newcommand{\vk}{\vec k}

\newcommand{\del}{\partial}

\begin{document}
	
	\title{Electrical conductivity of strongly magnetized dense quark matter - possibility of quantum hall effect}
	\author{Jayanta Dey}
	\affiliation{Indian Institute of Technology Bhilai, GEC Campus, Sejbahar, Raipur 492015, 
		Chhattisgarh, India}
		
	\author{Aritra Bandyopadhyay}
 \affiliation{Guangdong Provincial Key Laboratory of Nuclear Science, Institute of Quantum Matter, South China Normal University, Guangzhou 510006, China}
 \affiliation{Guangdong-Hong Kong Joint Laboratory of Quantum Matter, Southern Nuclear Science Computing Center, South China Normal University, Guangzhou 510006, China}
	
	\author{Akash Gupta}
	\affiliation{Ludwig-Maximilians-Universitat,  Theresienstra$\beta$e  37,  80333  Munchen,  Germany}
	
	\author{Naman Pujari}
	\affiliation{Indian Institute of Technology Bhilai, GEC Campus, Sejbahar, Raipur 492015, Chhattisgarh, India}
	
	\author{Sabyasachi Ghosh}
	\affiliation{Indian Institute of Technology Bhilai, GEC Campus, Sejbahar, Raipur 492015, 
		Chhattisgarh, India}

	\begin{abstract}
		\noindent
		We have pointed out the possibility of quantum Hall effect or quantum patterns of transportation in a degenerate strongly magnetized quark matter, which might be expected inside a highly dense compact star. An anisotropic pattern of electrical conductivity and resisitivity tensor in classical and quantum cases is explored by considering cyclotron motion and Landau quantization respectively. With increasing magnetic field, classical to quantum transitions are realized through enhanced/reduced resistivity/conductivity with jumping pattern. Considering QCD relaxation time scale of 10 fm, $eB\approx (1-4) m_\pi^2$ might be considered as strong magnetic field for massless and degenerate quark matter with quark chemical potential $\mu\approx 0.2-0.4$ GeV. Beyond these threshold ranges of magnetic field, perpendicular motion of quarks might be stopped and 3 $\rightarrow$ 1 dimensionally reduced conduction picture might be established.    
	\end{abstract}
	
	\maketitle
	
	\section{Introduction}
	\label{sec1}
	Compact stars like white dwarf (WD) and neutron stars (NS) are very unique
	astro-physical objects, which have long been attracted the attention of the theoretical physicists, especially due to their extremely high densities and presence of strong magnetic fields~\cite{Lai_rev}. The range of surface magnetic fields in NS is measured to be within $10^{12}$G~\cite{12B} to $10^{15}$G~\cite{15B} through various studies. Understanding the nature of the transport coefficients of this nuclear matter under extreme conditions with high density and magnetic field is considered to be crucial, as those play central roles in the astrophysical description of these compact stars. In the present work, among the transport coefficients we will focus on the electrical conductivity which is important while studying the decay of the magnetic field in the interiors of compact stars~\cite{Decay}. Simulations of magnetized neutron stars in the general relativity, including binary magnetized neutron star mergers involve solving relativistic magnetohydrodynamics equations with the electrical conductivity of the crustal matter as an input~\cite{MHD1,MHD2,MHD3,MHD4,Arus}. The binary neutron star merger simulation suddenly get special attention due to the recently observed gravitational wave signal GW170817~\cite{GW1}, which opened a new research field - multimessenger astronomy. The realistic picture of binary star merger is quite rigorous and the input quantity - the electrical conductivity might also be modified in time with the evolution. So microscopic calculation of electrical conductivity in presence of magnetic field might be an important research topic, connected with these contemporary research of multimessenger astronomy and gravitational wave. In absence of magnetic field, microscopic calculation of electrical conductivity for compact star has a long history with a long list of references (e.g. few selective Refs.~\cite{1950,1964,1970,1976}). In presence of magnetic field, Refs.~\cite{cond_B69,cond_B70,cond_B92,LL,Kerbikov,Baiko,Sedrakian_el} have estimated anisotropic conductivity of NS, whose quantum effect or Landau quantization has been explored by Refs.~\cite{LL,Kerbikov}. In one hand, the microscopic estimations of the electrical conductivity in presence of the magnetic field are important for compact stars, expected in high density and low temperature domain of the quantum chromodynamic (QCD) phase diagram. On the other hand, similar kind of microscopic calculations\cite{Nam,Hattori1,Manu1,Manu2,Feng_cond,Fukushima_cond,Arpan1,Arpan2,Arpan3,Dey1,Dey2,Asutosh,NJLB_el,SS_QM} become important for matter produced in heavy ion collision (HIC) experiments like relativistic heavy ion collision (RHIC) and large hadron collider (LHC), where high temperature and low density QCD phase diagram is expected. By increasing the temperature, hadron to quark phase transition is expected at nearly zero (net) quark/baryon density in RHIC or LHC experiments, where a huge magnetic field can be created in the peripheral collisions. Keeping that in mind, electrical conductivity of quark and hadronic matter in strong magnetic field is estimated by Refs.~\cite{Nam,Hattori1,Manu1,Manu2,Feng_cond,Fukushima_cond,Arpan1,Arpan2,Arpan3,Dey1,Dey2,Asutosh,NJLB_el,SS_QM}, among which Ref.~\cite{Dey1} has shown a very simple analytic expressions of different conductivity components for massless quark matter at high temperature and zero (net) quark/baryon density.
	In the present work, we have explored the estimations in the opposite domain, i.e. analytic expressions of different conductivity components for massless quark matter at zero temperature and high (net) quark/baryon density, expected in neutron star (NS) environment~\cite{Hatsuda,Kurkela}. Here we have explored the classical to the quantum transition of the conductivity tensor in a magnetic field axis, in which we have found changes from continuous to the quantized pattern for different components of conductivity. It is well known that due to the Lorentz force, $B=0$ isotropic conductivity breaks up into parallel, perpendicular and hall components in presence of the magnetic field. On the other hand, the quantum Hall effect (QHE) as well as quantum structures of other components are expected due to imposition of the Landau quantization, which are visible in a strong magnetic field domain only. Here, we have explored the possibility of QHE in (massless) quark matter, expected in the core of the NS~\cite{Hatsuda,Kurkela}.

	

	The paper is organized as follows. Next, Sec.~(\ref{sec2}) has gone through a quick reviewing the relaxation time approximation (RTA) based kinetic theory framework of  electrical conductivity in presence of magnetic field, whose finite temperature to degenerate case and classical to quantum modifications are done in two subsections. Their detail mathematical steps are given in appendix for completeness. Then in Sec.~(\ref{sec3}), we have numerically sketched the electrical conductivity and resistivity tensor components of (massless) quark matter, where the transition from classical to quantum aspects are discussed and the strong field domain for QHE in quark matter is explored. At the end, we have summarized our study in Sec.~(\ref{sec4}).

	\section{Formalism}
	\label{sec2}
	\subsection{Electrical conductivity in presence of a magnetic field}
	Let us assume an external magnetic field ${\vec B}=B {\hat z}$ is applied on the medium of relativistic quark matter with energy $\om$, 3-momentum $\vk$, mass $m$ and charge $e_Q$. When we apply electric filed ($E$) along x direction, we will get electrical dissipative flow along that direction with conductivity $\sigma_{xx}$. It will be reduced in presence of magnetic field because some part of flow will be deviated from x-direction and therefore, a non-zero Hall conductivity $\sigma_{xy}$ will be appeared into the picture. On the other hand, the conductivity along z-direction $\sigma_{zz}$ remain independent of magnetic field as Lorentz force couldn't affect the direction of the magnetic field. 
	Now, using relativistic Boltzmann's equation and with the help of relaxation time approximation (RTA) we can get the expression of $xx$, $xy$ and $zz$ components of conductivity as~\cite{Dey1,Dey2,Sedrakian_el},
	\bea
	\sigma_{xx}=\sigma_{yy}&=& \sigma_{xx}^Q+\sigma_{xx}^{\bar Q}~;~
	\sigma_{xy}=-\sigma_{yx}= \sigma_{xy}^Q+\sigma_{xy}^{\bar Q}~;~
	\sigma_{zz}= \sigma_{zz}^Q+\sigma_{zz}^{\bar Q}
	\nn\\
	{\rm where,}~
	\sigma_{xx}^{Q,\bar Q}&=&2g \beta \sum_{Q,\bar Q} e_{Q,\bar Q}^2 \int \frac{d^3k}{(2\pi)^3}
	\tau_c^{Q,\bar Q}\frac{1}{1+(\tau_c^{Q,\bar Q}/\tau_B^{Q,\bar Q})^2}\frac{k_x^2}{\om^2} f_0^{Q,\bar Q}(1- f_0^{Q,\bar Q})~,
	\label{sig_xxCl}
	\\
	\sigma_{xy}^{Q,\bar Q}&=&2g\beta \sum_{Q,\bar Q} e_{Q,\bar Q}^2 \int \frac{d^3k}{(2\pi)^3}
	\tau_c^{Q,\bar Q}\frac{\tau_c^{Q,\bar Q}/\tau_B^{Q,\bar Q}}{1+(\tau_c^{Q,\bar Q}/\tau_B^{Q,\bar Q})^2}\frac{k_y^2}{\om^2} f_0^{Q,\bar Q}(1- f_0^{Q,\bar Q})~,
	\label{sig_xyCl}
	\\
	\sigma_{zz}^{Q,\bar Q}&=&2g\beta \sum_{Q,\bar Q} e_{Q,\bar Q}^2 \int \frac{d^3k}{(2\pi)^3}
	\tau_c^{Q,\bar Q}\frac{k_z^2}{\om^2} f_0^{Q,\bar Q}(1- f_0^{Q,\bar Q})~.
	\label{sig_zzCl}
	\eea
	Detailed calculation yielding upto these equations is given in the appendices \ref{appA} and \ref{appB} for a relativistic electron gas.
	The components, Eq.~(\ref{sig_xxCl},\ref{sig_xyCl},\ref{sig_zzCl}) can also be respectively called as perpendicular ($\perp$), Hall ($\times$) and parallel ($\parallel$) components of conductivity, with reference to the magnetic field direction. Here, apart from spin degeneracy 2, $g$ counts for color degeneracy which is $3$; $Q$ and $\bar Q$ written in the script of
	quantities are for their particle and antiparticle contribution 
	respectively where we
	have consider only up $(u)$ and down $(d)$ flavor quarks.
	$e_{Q,\bar Q}$ represent their charge ($e_{u,\bar u}=\pm 2/3e,~ e_{d,\bar d}=\mp 1/3e,$); $f_0^{Q,\bar Q}=1\Big/\{e^{\beta(\om \mp \mu)}+1\}$ is Fermi-Dirac (FD) distribution function; $\tau_c^{Q,\bar Q}$ is thermal relaxation time; $\tau_B^{Q,\bar Q}=\frac{\om}{e_{Q,\bar Q}B}$ can be consider as magnetic relaxation time or cyclotron time period.
	For an average value of $\tau_B^{Q,\bar Q}$ as $\tau_B=\sum_Q \om_{\rm av}/(|e_{Q}|B)$ with  $\om_{\rm av}=\frac{\sum_{Q,\bar Q}\int\frac{d^3 k}{(2\pi)^3}\om f_0^{Q,\bar Q}}
	{\sum_{Q,\bar Q}\int \frac{d^3 k}{(2\pi)^3} f_0^{Q,\bar Q}}$, and at a constant
	value of $\tau_c^{Q,\bar Q}$ (say, $\tau_c$), conductivity components can be written in simple analytic form:
	\bea
	&&\sigma_{xx}=\sigma_{yy}=\sum_{Q} \sigma^Q_{D}\frac{1}{1+(\tau_c/\tau_B^Q)^2}
	+\sum_{\bar Q} \sigma^{\bar Q}_{D}\frac{1}{1+(\tau_c/\tau_B^{\bar Q})^2}~,
	\nn\\
	&&\sigma_{yx}=-\sigma_{xy}=\sum_{Q} \sigma^Q_{D}\frac{(\tau_c/\tau_B^{\bar Q})}{1+(\tau_c/\tau_B^Q)^2}
	+\sum_{\bar Q} \sigma^{\bar Q}_{D}\frac{(\tau_c/\tau_B^{\bar Q})}{1+(\tau_c/\tau_B^{\bar Q})^2}~,
	\nn\\
	&&\sigma_{zz}=\sigma_{D}^Q + \sigma_{D}^{\bar Q}
	\eea
	where,
	$\sigma_{D}^{Q,\bar Q} = 2g \beta \sum_{Q,\bar Q} e_{Q,\bar Q}^2 \int \frac{d^3k}{(2\pi)^3}\tau_c^{Q,\bar Q}\frac{k^2}{3\om^2} f_0^{Q,\bar Q}(1- f_0^{Q,\bar Q})$
	can be considered as relativistic form of Drude conductivity,
	whose non-relativistic form $\sigma_D=\frac{ne^2\tau_c}{m}$ for net
	density $n=2g\sum_{Q,\bar Q} \int \frac{d^3 k}{(2\pi)^3} (f_0^Q-f_0^{\bar{Q}})$ is well known to us. 
	
	Further, the elements of resistivity matrix can also be written in terms of 
	the core component of the conductivity, i.e. $\sigma_{D}$ as (Appendix \ref{appB})
	\bea
	&&\rho_{xx}=\rho_{yy}=\frac{1}{
		\sum_Q\frac{1}{\rho_{xx}^Q} + \sum_{\bar Q}\frac{1}{\rho_{xx}^{\bar Q}}
	} = \frac{1}{ \sum_Q\sigma_{D}^Q + \sum_{\bar Q}\sigma_{D}^{\bar Q} }
	=\frac{1}{\sigma_{zz}} = \rho_{zz}
	\nn\\
	&&\rho_{xy}=-\rho_{yx}= \frac{1}{
		\sum_Q\frac{1}{\rho_{xy}^Q} + \sum_{\bar Q}\frac{1}{\rho_{xy}^{\bar Q}}
	}
	=\frac{1}{
		\sum_Q\sigma_D^Q\frac{\tau_{B}^Q}{\tau_c} + \sum_{\bar Q}\sigma_D^{\bar Q}\frac{\tau_{B}^{\bar Q}}{\tau_c}	}~.
	\label{xy_D}
	\eea
	One can get the standard expression of (classical) Hall resistivity~\cite{Tong}
	$\rho_{xy}=\frac{B}{ne}$, when we use non-relativistic Drude conductivity $\sigma_{D}=\frac{ne^2\tau_c}{m}$
	and cyclotron frequency $(\tau_B)^{-1}=\frac{eB}{m}$. 
	%
	
	%
	At $T=0$,  if we analyze FD distribution function numerically then
one can identify that decreasing temperature $T$ and increasing chemical potential $\mu$ will push the FD distribution towards a step function. So one can consider degenerate or $T=0$ picture for high density and low temperature environment like neutron star. Hence, we can modify our earlier calculations for degenerate picture, where finite temperature FD distribution function $f_0^Q$ will be replaced by step function $\theta(\mu -\om)$. Note that the antiparticles give null contribution at $T=0$, so only particle contribution will count. Now, number density for the degenerate gas at $T=0$ will be,
	\bea
	n&=&2g \sum_{Q} \int \frac{d^3k}{(2\pi)^3} \theta(\mu-\om)
	\nn\\
	&=&{\frac{4g}{6\pi^2}} k_F^3~,~{\rm where}~ \mu=\{k_F^2+m^2\}^{1/2}~.
	\label{n_den}
	\eea
	During the electrical conductivity calculation for $T=0$, only we have to use the replacement:
	\be
	\frac{\del f_0^Q}{\del \om}=-\beta f^Q_0(1-f^Q_0)
	\rightarrow 
	\frac{\del}{\del \om}\theta(\mu-\om)=-\delta(\om-\mu)~.
	\ee
	So the classical expressions, given in Eqs.~(\ref{xy_D}), are modified to
	\begin{subequations}
	\label{xy_F}
	\bea
	&&\sigma_{xx}=\sum_{Q=u,d} \sigma_{xx}^Q=\sum_{Q=u,d} \sigma_F^Q\frac{1}{1+(\tau_c/\tau_{BF}^Q)^2}=\sigma_{yy}
	\\
	&&\sigma_{yx}=\sum_{Q=u,d}\sigma_{yx}^Q=\sum_{Q=u,d}\sigma_F^Q \frac{\tau_c/\tau_{BF}^Q}{1+(\tau_c/\tau_{BF}^Q)^2}
	=-\sigma_{xy}
	\\
	&& \sigma_{zz}=\sum_{Q=u,d}\sigma_{zz}^Q=\sum_{Q=u,d}\sigma_F^Q
	\\
	&&\rho_{xx}=\frac{1}{\sum_Q\frac{1}{\rho_{xx}^Q}}=\frac{1}{\sum_Q\sigma_{F}^Q}=\rho_{yy}=\rho_{zz}
	\\
	&&\rho_{xy}=\frac{1}{\sum_Q\frac{1}{\rho_{xy}^Q}}=\frac{1}{\sum_Q\sigma_F^Q\frac{\tau_{BF}^Q}{\tau_c}}=-\rho_{yx}~,
	\eea
	\end{subequations}
	where $\sigma_F$ is the degenerate/$T=0$ version of $\sigma_D$, given as
	\bea
	\sigma_F^Q&=& \sum_Q 2g e_Q^2 \int \frac{d^3k}{(2\pi)^3}
	\tau_c\frac{\vk^2}{3\om^2} \delta(\om -\mu)
	\nn\\
	&=& \sum_Q 2g \frac{e_Q^2}{6\pi^2}\tau_c\frac{(\mu^2-m^2)^{3/2}}{\mu}
	\Theta(\mu^2-m^2)~.
	\label{sF_cl}
	\eea

	\subsection{Quantum version expression of electrical conductivity}
	So far we have not considered the Landau quantizations - a quantum aspects of external 
	magnetic field. 
	In quantum picture, the energy $\om= (\vk^2+m^2)^{1/2}$ and 
	phase space $2\int \frac{d^3k}{(2\pi)^3}$ will be modified as $\om_{l} = (k_z^2+m^2+2l|{e_Q}|B)^{1/2}$
	and $\sum_{l=0}^\infty \alpha_l \frac{|{ e}|B}{2\pi} 
	\int\limits^{+\infty}_{-\infty} \frac{dk_z}{2\pi}$~,
	where spin degeneracy 2 will be converted to $\alpha_l=2-\delta_{0,l}$,
	which will be 1 for lowest Lnadau level (LLL) $l=0$ and 2 for remaining levels $l$. 
	We can also roughly assume, $k_x^2\approx k_y^2\approx (\frac{k_x^2+k_y^2}{2})=\frac{2l{\tilde e}B}{2}$.
	Adopting these quantum impositions, the expressions of conductivity
	will be changed to
	\bea
	\sigma_{xx}^{Q,\bar Q} = g \beta \sum_Q e_Q^2  \sum_{l=0}^\infty \alpha_l \frac{|{e_Q}|B}{2\pi} 
	\int\limits^{+\infty}_{-\infty} \frac{dk_z}{2\pi} \frac{{l|{ e_Q}|B}}{\om^2_{l}} \tau_c^Q \frac{1}{1+(\tau_c^Q/\tau_B^Q)^2}f_0^Q(\om_l)[1-f_0^Q(\om_l)]
	\nn\\
	\sigma_{xy}^{Q,\bar Q} = g \beta \sum_Q e_Q^2  \sum_{l=0}^\infty \alpha_l \frac{|{ e_Q}|B}{2\pi} 
	\int\limits^{+\infty}_{-\infty} \frac{dk_z}{2\pi} \frac{{l|{e_Q}|B}}{\om^2_{l}} \tau_c^Q \frac{\tau_c^Q/\tau_B^Q}{1+(\tau_c^Q/\tau_B^Q)^2}f_0^Q(\om_l)[1-f_0^Q(\om_l)]
	\nn\\
	\sigma_{zz}^{Q,\bar Q} = g \beta \sum_Q e_Q^2  \sum_{l=0}^\infty \alpha_l \frac{|{e_Q}|B}{2\pi} 
	\int\limits^{+\infty}_{-\infty} \frac{dk_z}{2\pi} \frac{k^2_z}{\om^2_{l}} \tau_c^Q 
	f_0^Q(\om_l)[1-f_0^Q(\om_l)]~.
	\label{Lsig_QM}
	\eea
	If we compare Eq.~(\ref{Lsig_QM}) with Eq.~(\ref{xy_D}), we can identify quantum
	version of $\sigma_D$,
	\be
	\sigma^{Q{\langle\rm QM\rangle}}_{D\perp}= \sum_{Q=u,d} g \beta e_Q^2 \sum_{l=0}^\infty \alpha_l \frac{l(|e_Q|B)^2}{(2\pi)^2} 
	\int\limits^{\infty}_{-\infty} dk_z \frac{1}{\om^2_l} \tau_c f_0^Q(\om_l)[1-f_0^Q(\om_l)]~,
	\ee
which is true for all perpendicular ($\perp$) components: $xx, yy ~{\rm and}~ xy$. For the $zz$ component of conductivity which is parallel ($\parallel$) to the magnetic field, quantum version of $\sigma_{D}$ would be
	\be
\sigma^{Q{\langle\rm QM\rangle}}_{D\parallel}= \sum_{Q=u,d} g \beta e_Q^2 \sum_{l=0}^\infty \alpha_l \frac{(|e_Q|B)}{(2\pi)^2} 
\int\limits^{\infty}_{-\infty} dk_z \frac{k_z^2}{\om^2_l} \tau_c f_0^Q(\om_l)[1-f_0^Q(\om_l)]~.
\ee 
	
	Furthermore, when we go to the quantum version of the degenerate/$T=0$ case, $\sigma_F$ gets modified. For the $xx, yy ~{\rm and}~ xy$ components it can be expressed as 
	\bea
	\sigma^{Q{\langle\rm QM\rangle}}_{F\perp}&=& \sum_{Q=u,d}g {e_Q}^2  \sum_{l=0}^\infty \alpha_l \frac{l(|e_Q|B)^2}{(2\pi)^2} 
	\int\limits^{\infty}_{-\infty} dk_z \frac{1}{\om^2_l} \tau_c \delta(\om_l -\mu)
	\nn\\
	&=& \sum_{Q=u,d} g {e_Q}^2  \sum_{l=0}^\infty \alpha_l \frac{l(|e_Q|B)^2}{2\pi^2} 
	\frac{\tau_c}{\mu(\mu^2-2le_QB-m^2)^{1/2}}~\Theta(\mu^2-2le_QB-m^2)~,
	\label{sD_QM_T0}
	\eea
	and for the $zz$ component
	\bea
	\sigma^{Q{\langle\rm QM\rangle}}_{F\parallel}&=& \sum_{Q=u,d} g {e_Q}^2  \sum_{l=0}^\infty \alpha_l \frac{(|e_Q|B)}{(2\pi)^2} 
	\int\limits^{\infty}_{-\infty} dk_z \frac{k_z^2}{\om^2_l} \tau_c \delta(\om_l -\mu)
	\nn\\
	&=& \sum_{Q=u,d} g {e_Q}^2  \sum_{l=0}^\infty \alpha_l \frac{(|e_Q|B)}{2\pi^2} 
	\frac{\tau_c (\mu^2-2le_QB-m^2)^{1/2}}{\mu}~\Theta(\mu^2-2le_QB-m^2)~.
	\label{sD_QM_T1}
	\eea
	
	Here, the kinematical constrain, set by the Heaviside function $\Theta(\mu^2-2le_QB-m^2)$ in Eqs.~(\ref{sD_QM_T0}) and (\ref{sD_QM_T1}) implies that the sum over Landau levels is restricted up to a maximum (integer) value $l^Q_{\rm max}$, which is a function of $e_Q$ and can be obtained as 
	\be 
	l^Q_{\rm max} = {\rm Integer}\left[\frac{\mu^2-m^2}{2e_QB}\right].
	\label{l_max}
	\ee
	
	Emergence of $l^Q_{\rm max}$ can also be explained from the degenerate version of the quantized number density, i.e. Eq.~(\ref{n_den}):
	\bea
	n^{\rm QM}&=&\sum_{Q} \sum^\infty_{l=0} \frac{|e_Q|B}{(2\pi)^2}\int_{-\infty}^{\infty} dk_z \theta (\mu-\om_l)
	\nn\\
	&=&\sum_{Q}\sum^\infty_{l=0} \frac{|e_Q|B}{2\pi^2}(\mu^2-2le_QB-m^2)^{1/2}
	\nn\\
	&=&\sum_{Q}\sum^\infty_{l=0} \frac{|e_Q|B}{2\pi^2}k_z^F(l)~.
	\label{nQMF}
	\eea
	The Fermi momentum along $z$-axis, $k_z^F$ will depend on Landau level $l$ and they are related with the Fermi energy or chemical potential $\mu$ as
	\bea
	\mu^2 &=& [k_z^F(l=0)]^2+m^2
	\nn\\
	&=& [k_z^F(l=1)]^2+2e_QB+m^2
	\nn\\
	&=& [k_z^F(l=2)]^2+4e_QB+m^2
	\nn\\
	&=& ...
	\nn\\
	&=& ...
	\nn\\
	&=& [k_z^F(l=l^Q_{\rm max}-1)]^2+2(l^Q_{\rm max}-1)e_QB+m^2
	\nn\\
	&=& 2l^Q_{\rm max}e_QB+m^2~.
	\eea
	We see that as Landau level $l$ increases, $k_z^F$ will decrease.
	So, for lowest Landau level (LLL), we will get highest 
	$k_z^F=\{\mu^2-m^2\}^{1/2}$, which is basically Fermi momentum $k_F$
	of classical picture. On the other hand, for maximum value of Landau level
	$l^Q_{\rm max}$, we can expect $k_z^F(l=l^Q_{\rm max})\approx 0$. So, we can get a similar relation for $l^Q_{\rm max}$ as in Eq.~(\ref{l_max}),
	\bea
	2l^Q_{\rm max}e_QB&=&[k_z^F(l=0)]^2=\mu^2-m^2
	\nn\\
	\Rightarrow l^Q_{\rm max} &=& {\rm Integer}\left[\frac{k_F^2}{2e_QB}\right]={\rm Integer}\left[\frac{\mu^2-m^2}{2e_QB}\right]~.
	\eea
	
	So, now with the inclusion of $l^Q_{\rm max}$ we can further simplify the expression of $\sigma^{Q{\langle\rm QM\rangle}}_{F\perp}$ as - 
	\bea
	\sigma^{Q{\langle\rm QM\rangle}}_{F\perp}&=& \sum_{Q=u,d}g e_Q^2  \sum_{l=0}^{l^Q_{\rm max}-1} \alpha_l \frac{l(|e_Q|B)^2}{(2\pi)^2} \frac{ \tau_c}{\mu(\mu^2-2le_QB-m^2)^{1/2}}~, \nn\\
	&=& \sum_{Q=u,d}g e_Q^2\tau_c\frac{2(|e_Q|B)^2}{2\pi^2\mu\sqrt{2e_QB}}   \sum_{l=1}^{l^Q_{\rm max}-1} \frac{l}{(\frac{\mu^2-m^2}{2e_QB}-l)^{1/2}} ~,\nn\\
	&=& \sum_{Q=u,d}g e_Q^2\tau_c\frac{(|e_Q|B)^2}{\pi^2\mu\sqrt{2e_QB}}   \sum_{l=1}^{l^Q_{\rm max}-1} \frac{l}{(l^Q_{\rm max}-l)^{1/2}}.
	\label{sigmaF_exact}
	\eea
	The above mentioned sum can be regularized with the combination of Hurwitz Zeta function $\zeta(s,a)$ and Reimann Zeta function $\zeta(s)$~\cite{Elizalde:1994gf,Bandyopadhyay:2018gbw}, i.e., 
	\bea
	\zeta(s,a) &=& \sum_{l=0}^\infty \frac{1}{(l+a)^s}, \nn\\
	\zeta(s) &=& \sum_{l=1}^\infty \frac{1}{l^s}.
	\eea 
	Hence to apply the zeta function regularization, we break down the sum into two parts and subsequently write those parts in terms of zeta functions as 
	\bea
	&&\sum_{l=1}^{l^Q_{\rm max}-1} \frac{l}{(l^Q_{\rm max}-l)^{1/2}} =  l^Q_{\rm max}\sum_{l=1}^{l^Q_{\rm max}-1} \frac{1}{(l^Q_{\rm max}-l)^{1/2}}- \sum_{l=1}^{l^Q_{\rm max}-1} \frac{1}{(l^Q_{\rm max}-l)^{-1/2}},\nn\\
	&&=  l^Q_{\rm max}\left[\zeta\left(\frac{1}{2}\right)-\zeta \left(\frac{1}{2},l^Q_{\rm max}\right)\right]- \left[\zeta \left(-\frac{1}{2}\right) -\zeta \left(-\frac{1}{2},l^Q_{\rm max}\right)\right], \nn\\
	&&= \zeta \left(-\frac{1}{2},l^Q_{\rm max}\right) -l^Q_{\rm max}~ \zeta \left(\frac{1}{2},l^Q_{\rm max}\right) - \zeta \left(-\frac{1}{2}\right) + l^Q_{\rm max}~\zeta \left(\frac{1}{2}\right).
	\eea
	So the final expression for $\sigma^{Q{\langle\rm QM\rangle}}_{F\perp}$ comes out to be 
	\bea
	\sigma^{Q{\langle\rm QM\rangle}}_{F\perp} = \sum_{Q=u,d}\frac{g e_Q^2\tau_c^Q(|e_Q|B)^{\frac{3}{2}}}{\sqrt{2}\pi^2\mu} \left[\zeta \left(-\frac{1}{2},l^Q_{\rm max}\right) -l^Q_{\rm max}~ \zeta \left(\frac{1}{2},l^Q_{\rm max}\right) - \zeta \left(-\frac{1}{2}\right) + l^Q_{\rm max}~\zeta \left(\frac{1}{2}\right)\right].
	\label{sigmaF_zeta}
	\eea
	
Similarly, from Eq.~(\ref{sD_QM_T1}), $\sigma^{Q{\langle\rm QM\rangle}}_{F\parallel}$ would become
\bea
\sigma^{Q{\langle\rm QM\rangle}}_{F\parallel} &=& \sum_{Q=u,d}g{e_Q}^2  \sum_{l=0}^{l^Q_{\rm max}-1} \alpha_l \frac{(|e_Q|B)}{2\pi^2} \frac{\tau_c^Q (\mu^2-2le_QB-m^2)^{1/2}}{\mu}
\nn\\
&=&  \sum_{Q=u,d}\frac{g{e_Q}^2\tau_c^Q(|e_Q|B)^{\frac{3}{2}}}{\sqrt{2}\pi^2\mu}  \left( \sqrt{l^Q_{\rm max}} + 2\sum_{l=1}^{l^Q_{\rm max}-1} (l^Q_{\rm max}-l)^{1/2} \right), \nn\\
&=& \sum_{Q=u,d}\frac{g{e_Q}^2\tau_c^Q(|e_Q|B)^{\frac{3}{2}}}{\sqrt{2}\pi^2\mu}  \left[ \sqrt{l^Q_{\rm max}} + 2\zeta \left(-\frac{1}{2}\right) -2\zeta \left(-\frac{1}{2},l^Q_{\rm max}\right) \right].
\label{sigmaF_zeta_sp}
\eea 

Hence, finally following the classical case, the expressions of electrical conductivity and resistivity for quantum degenerate case can be written in terms of  $\sigma^{Q{\langle\rm QM\rangle}}_{F\perp}$ and $\sigma^{Q{\langle\rm QM\rangle}}_{F\parallel}$ as
\begin{subequations}
\label{final_exprsn_qm}
	\bea
	&&\sigma_{xx}^{\rm QM}=\sum_{Q=u,d} \sigma^{Q{\langle\rm QM\rangle}}_{F\perp}\frac{1}{1+(\tau_c/\tau_{BF}^Q)^2}
	=\sigma_{yy}^{\rm QM}
	\\
	&&\sigma_{yx}^{\rm QM}=\sum_{Q=u,d}\sigma^{Q{\langle\rm QM\rangle}}_{F\perp} \frac{\tau_c/\tau_{BF}^Q}{1+(\tau_c/\tau_{BF}^Q)^2}
	=-\sigma_{xy}^{\rm QM}
	\\
	&& \sigma_{zz}^{\rm QM}=\sum_{Q=u,d}\sigma^{Q{\langle\rm QM\rangle}}_{F\parallel}
	\\
	&&\rho_{xx}^{\rm QM}=\frac{1}{\sum_Q\sigma^{Q{\langle\rm QM\rangle}}_{F\perp}}=\rho_{yy}^{\rm QM}
	\\
	&&\rho_{xy}^{\rm QM}=\frac{1}{\sum_Q\sigma^{Q{\langle\rm QM\rangle}}_{F\perp}\frac{\tau_{BF}^Q}{\tau_c}}=-\rho_{yx}^{\rm QM}
	\\
	&&\rho_{zz}^{\rm QM}=\frac{1}{\sum_Q\sigma^{Q{\langle\rm QM\rangle}}_{F\parallel}}~.
	\eea
	\end{subequations}

	\section{Results}
	\label{sec3}
	\begin{figure}  
		\centering	  
		\includegraphics[scale=0.4]{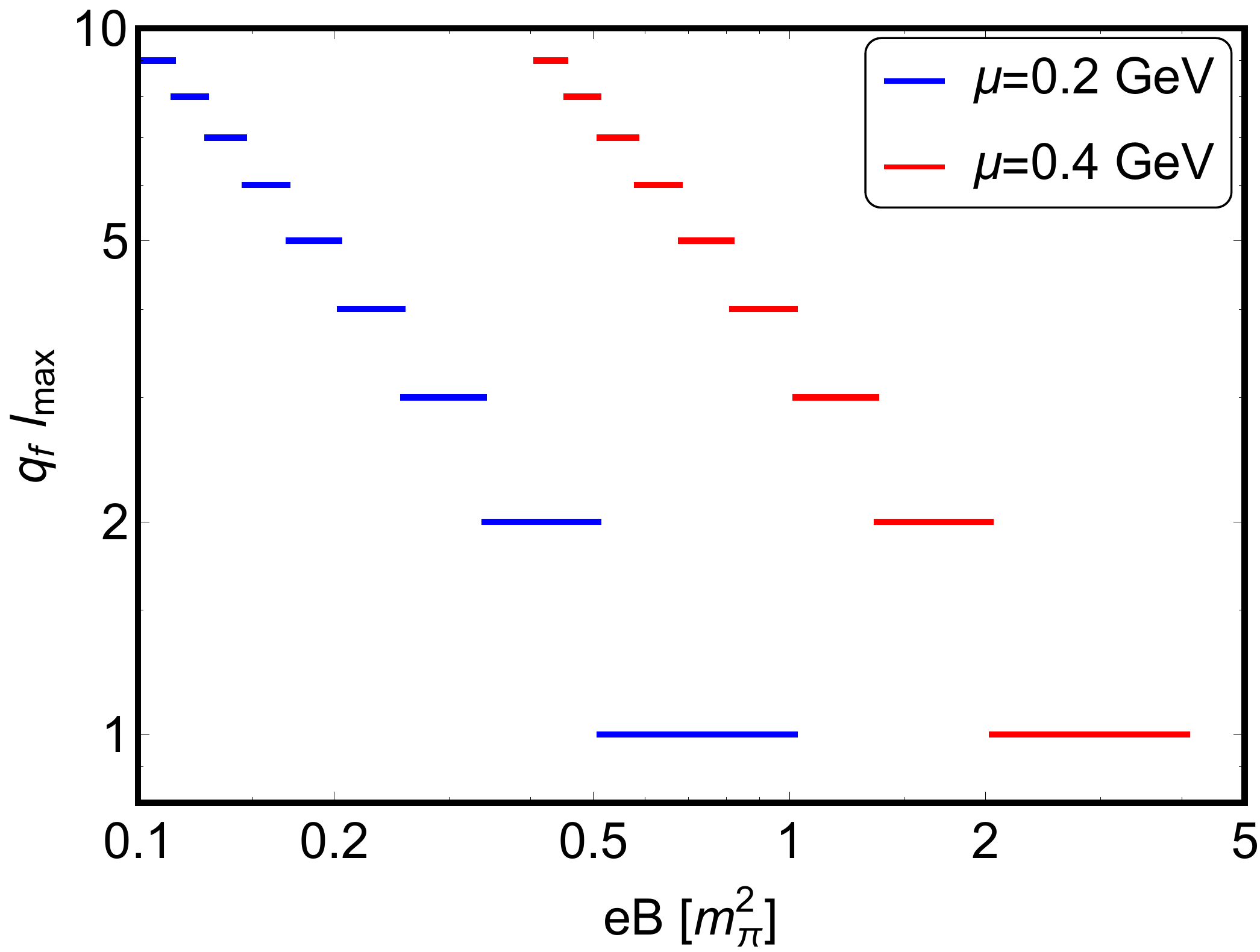} 
		\caption{Maximum Landau level $l^Q_{\rm max}$ as a function of magnetic field in quantized way for two values of chemical potential $\mu=0.2$ GeV, $\mu=0.4$ GeV. $q_f$ represents the normalization factor $e^Q/e$.}
		\label{lmax_B}
	\end{figure} 
	Here, we want to apply the RTA based kinetic theory framework, addressed in earlier section, for estimating electrical conductivity of massless and highly dense quark matter, facing strong magnetic field. The picture might be expected in the core of NS~\cite{Hatsuda,Kurkela}. In the formalism, though we have addressed expressions for both $T\neq 0$ and $T=0$ cases, but in the results our focus will be on the latter case only, i.e. the classical and quantum expressions for conductivity and resistivity components at $T=0$. It is well known that magnetic field creates an anisotropic flow in a charged fluid through Lorentz force~\cite{Landau,Dey1}. It means that the isotropic property of conductivity tensor $\sigma^{ij}\delta^{ij}\sigma$ (i.e. $\sigma^{xx}=\sigma^{yy}=\sigma^{zz}=\sigma$) at $B=0$ is destroyed when an external magnetic field is applied.   
	Based on the conception of classical electrodynamics, Lorentz force will modify the RTA based kinetic theory calculation, as discussed in the formalism section and we will get final expressions of conductivity tensor in Eqs.~(\ref{xy_F}), which might be called as classical expressions. 
	On the other hand, in quantum picture, it is well known that perpendicular momentum components are quantized, commonly known as Landau quantization. Incorporating that quantum aspects in RTA based expressions, we get their quantum expressions, given in Eqs.~(\ref{final_exprsn_qm}). Readers can notice that Landau levels go to infinity for $T\neq 0$ but it terminates to a maximum integer value $l^Q_{\rm max}(e_QB)$ for $T=0$ as the probability distribution is restricted within Fermi momentum range, giving a kinematical constrain to the system. To understand this picture in a better way, we have shown  Fig.~\ref{lmax_B}, where $l_{max}$ for two different values of chemical potential ($\mu$) are plotted against $B$-axis. The Eq.(\ref{l_max}) is used to find $l_{max}$ for different magnetic field at $\mu = 0.2,~0.4$ GeV, where magnetic field axis is taken in terms of pion mass square ($m_\pi^2$). As value of $l$ should be integer, so we kept $l_{max}$ fix to the previous integer value until it changes to next the integer with magnetic field. Classical to quantum transition effect as Landau quantization can easily be followed as continuum to discrete values of $l_{max}$. By increasing magnetic field, $l_{max}$ is gradually reduced and beyond a certain threshold $B_{0}$, until which $l_{max}=1$, we will get $l=0$, which wil thenl be extended from $B_0$ to infinity. For example, Fig.~\ref{lmax_B} suggests $B_0\approx m_\pi^2$, $4m_\pi^2$ for $\mu=0.2$, $0.4$ GeV respectively. This means that if we expect massless quark matter at high dense environment with $\mu=0.2$ or $0.4$ GeV, then for $B > m_\pi^2$ or $4m_\pi^2$, perpendicular motion of the medium will stop abruptly and 3D transport phenomena will be suddenly converted to 1D phenomena. In this situation, only the longitudinal conductivity $\sigma^{zz}$ will exist and its expression will become:
	\bea 
	\sigma^{zz}_{QM} &=& \sum_Q g {e_Q}^2 \frac{(|e_Q|B_0)}{2\pi^2} 
	\frac{(\mu^2-m^2)^{1/2}}{\mu} \tau_c^Q
	\label{szz_l0}
	\eea 
	which will be more simplified to
	\be
	\sigma^{zz}_{QM}= \sum_Q g {e_Q}^2 \frac{(|e_Q|B_0\tau_c^Q)}{2\pi^2} ~,
	\label{szz_lm0}
	\ee 
	for massless case.

	\begin{figure}  
		\centering	  
		\includegraphics[scale=0.32]{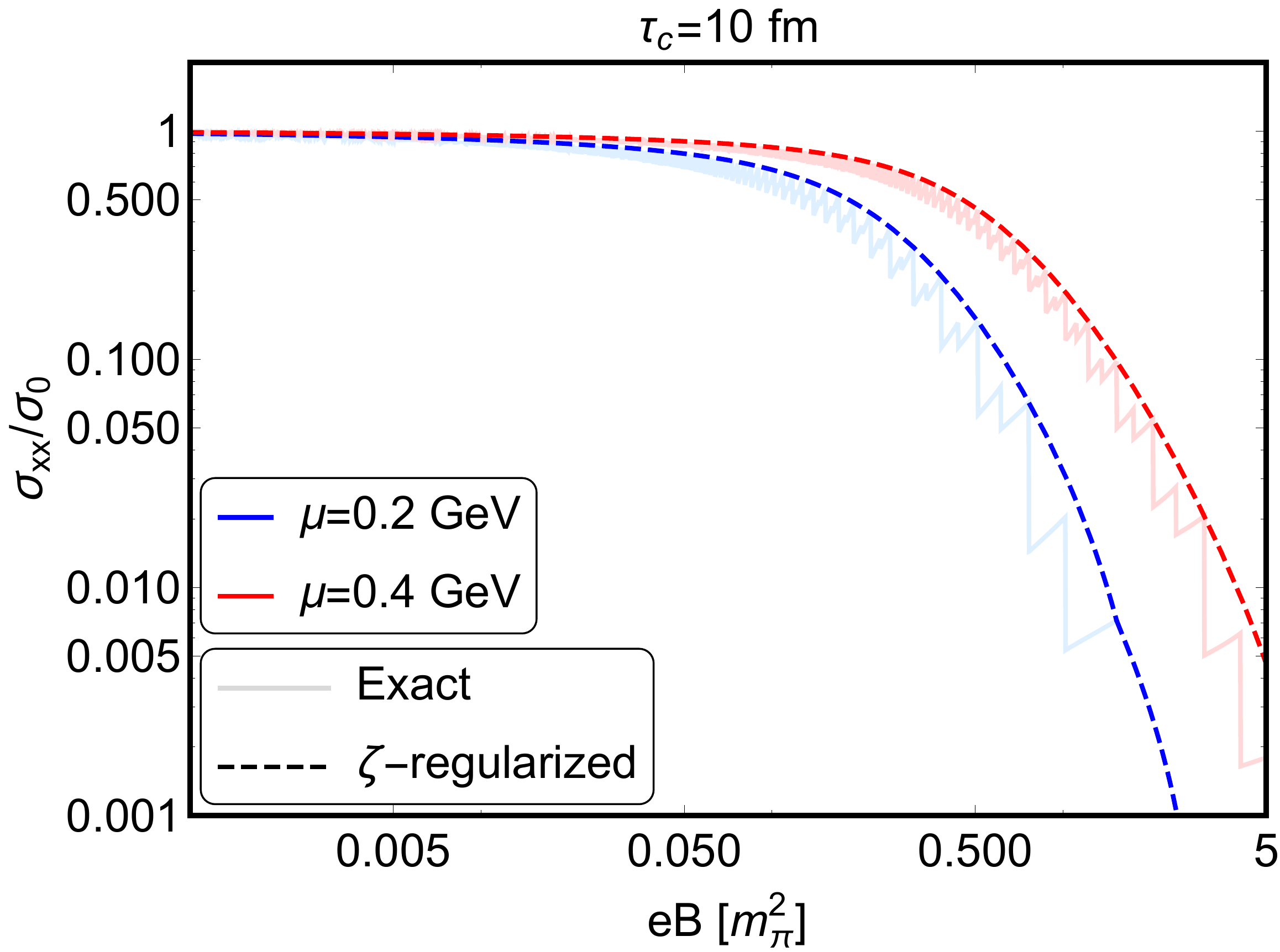} 
		\includegraphics[scale=0.32]{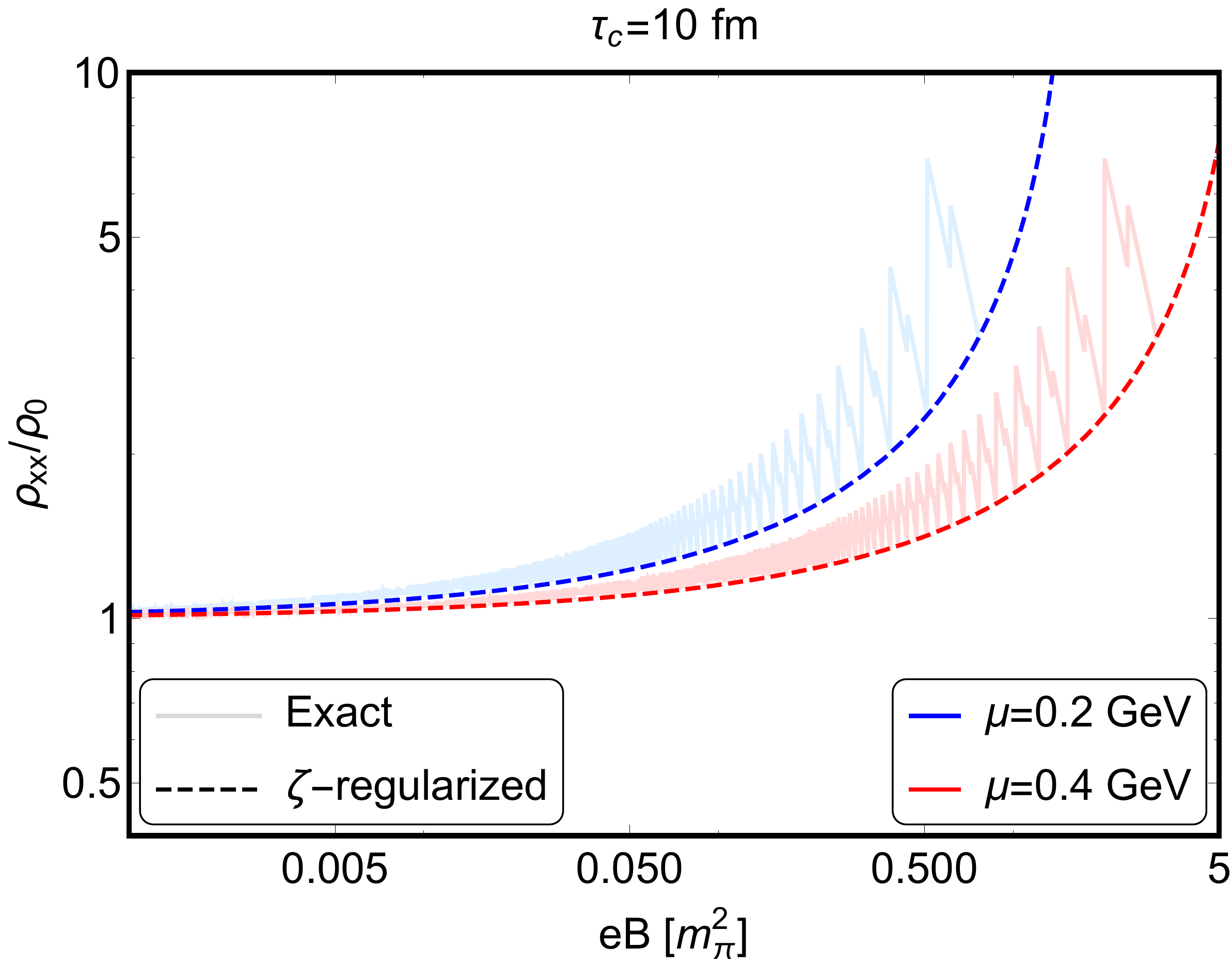}
		\includegraphics[scale=0.32]{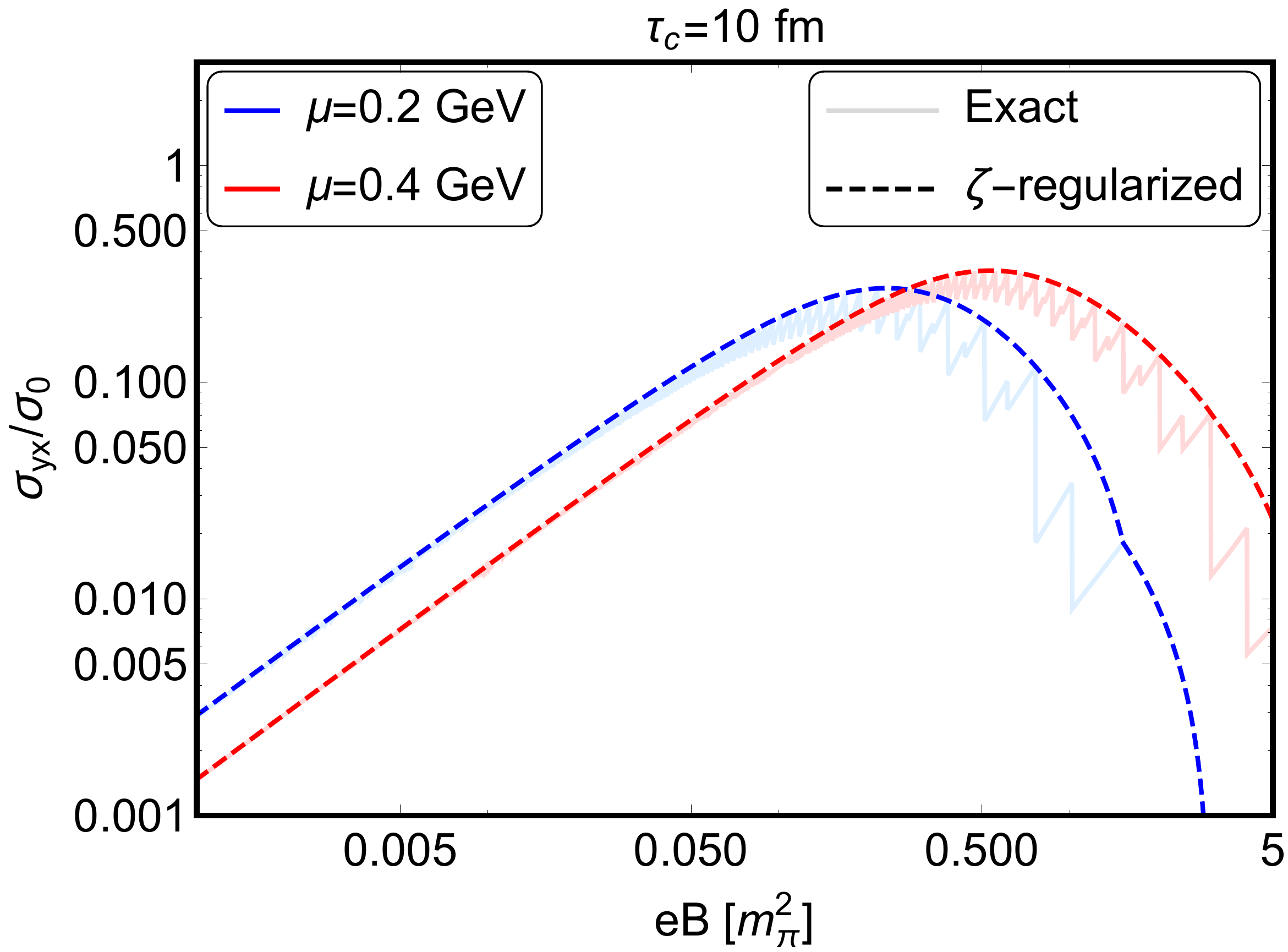}
		\includegraphics[scale=0.32]{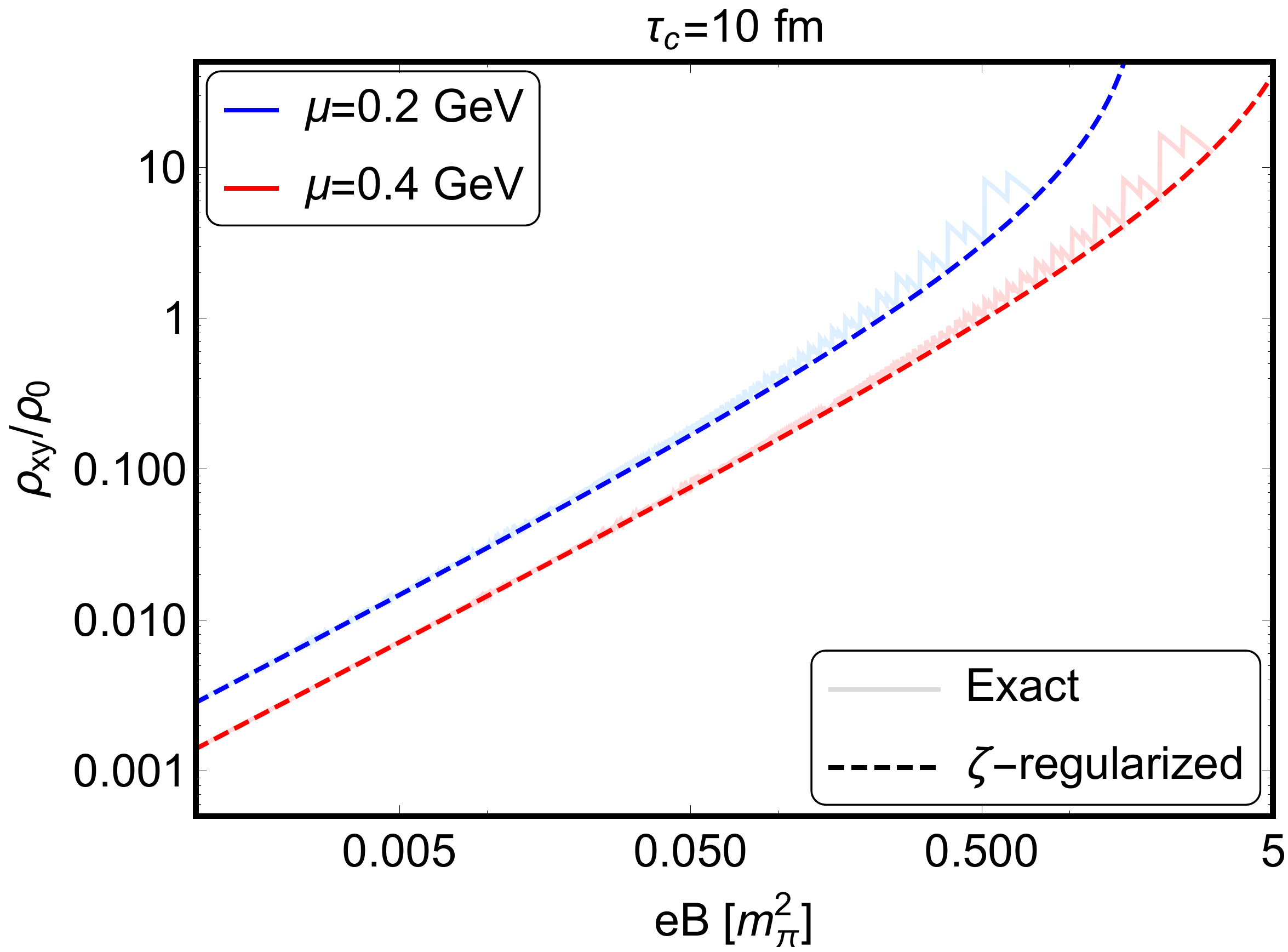}
		\includegraphics[scale=0.32]{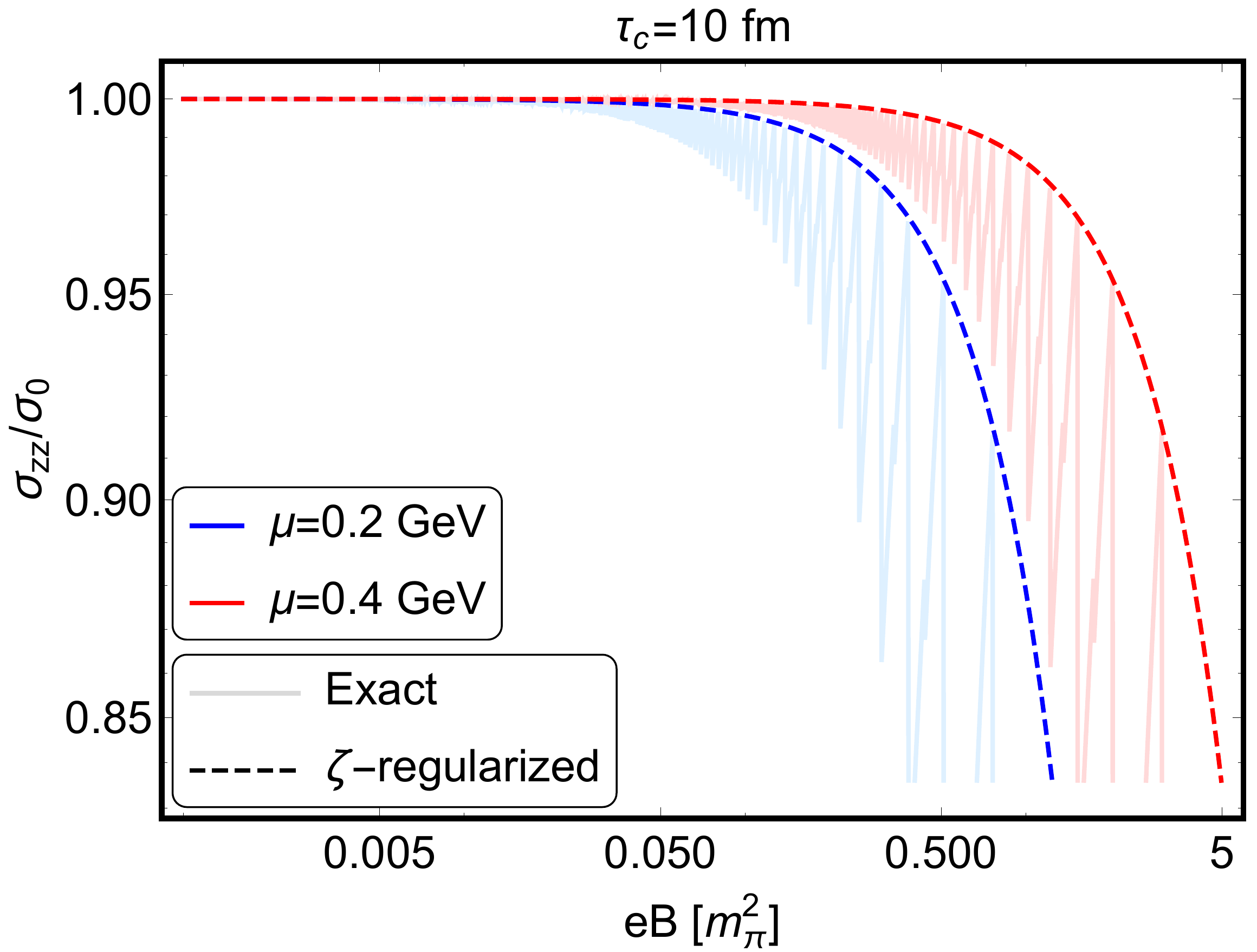}
		\includegraphics[scale=0.32]{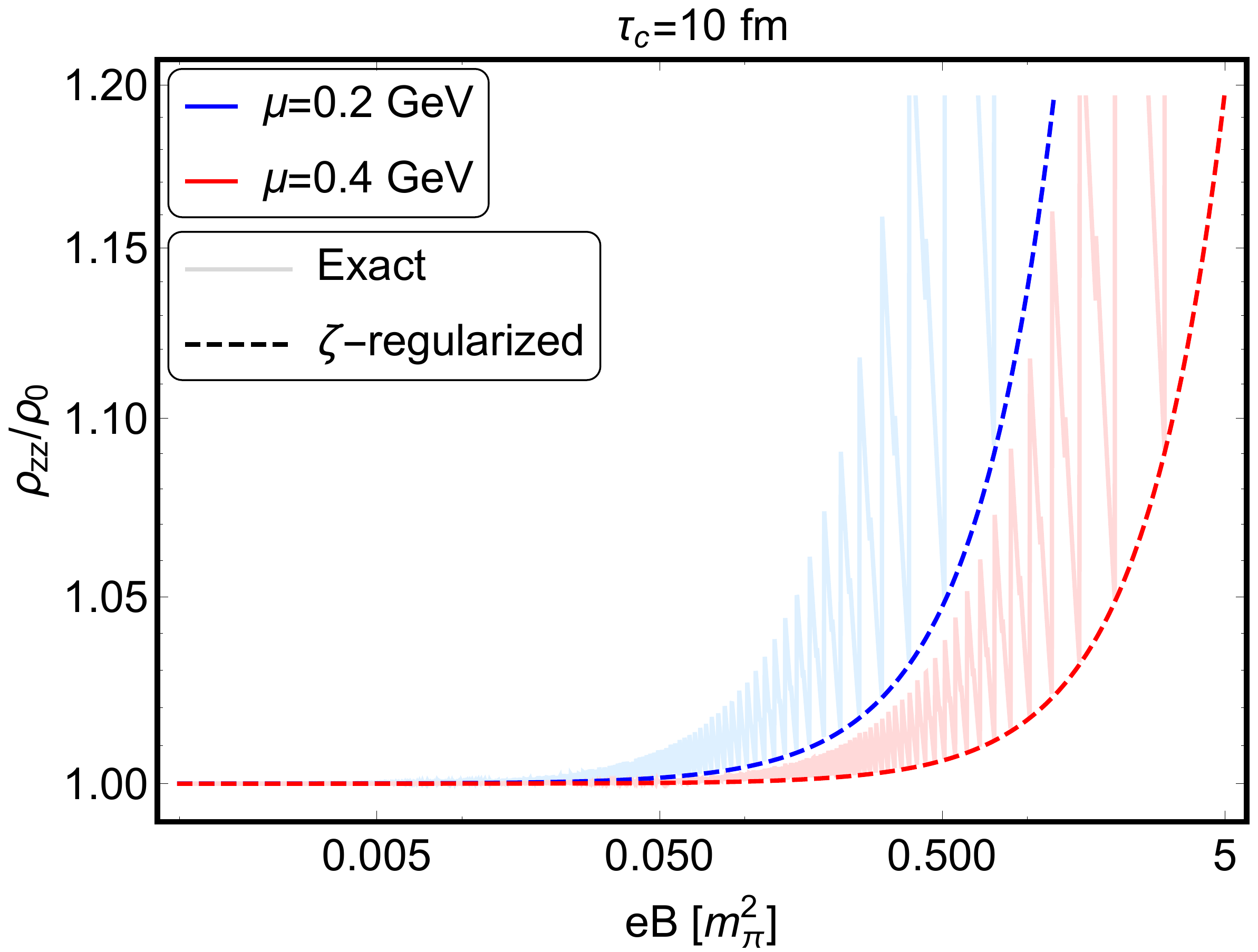}
		\caption{Scaled perpendicular, Hall and longitudinal conductivity ($\sigma_{xx}$, $\sigma_{xy}$ and $\sigma_{zz}$) and resistivity ($\rho_{xx}$, $\rho_{xy}$ and $\rho_{zz}$) vs magnetic field, normalized by $m_\pi^2$. $\sigma_0$ and $\rho_0$ respectively signify the conductivity and resistivity for $eB=0$. Exact estimations (fade solid line) are fluctuating due to Landau level summation, but their envelop or $\zeta$-regularized values (dash line) are continous in nature.}
		\label{sig_QM}
	\end{figure} 
	Understanding the $B$-dependence of $l_{max}$, next we have shown the conductivity and resistivity components in Fig.~(\ref{sig_QM}) by using the exact quantum expression, given in Eq.~(\ref{sigmaF_exact}) and also using the $\zeta$-regularized expression, given in Eqs.~(\ref{sigmaF_zeta}) and (\ref{sigmaF_zeta_sp}). One can find here the fluctuating numerical values of $\sigma_{xx, xy, zz}$ and $\rho_{xx, xy, zz}$ because of the discreet jumps within the Landau level summation. When we go from low to high $B$, the changes of pattern in fluctuation can be noticed, which are connected with the transformation from classical to quantum zones. If we focus on the envelop or $\zeta$-regularized values (dash line) of exact fluctuating numerics (fade solid line) in Fig.~(\ref{sig_QM}), then the interval of points of that envelop are quite distinguishable in high $B$ zone, which reflect quantum jump of conductivity or resistivity values. Those jumps for the Hall components of conductivity/resistivity is popularly known as the quantum Hall effect (QHE)~\cite{Tong}. This phenomenon is quite exotic as microscopic Landau quantization aspects of charge particle is reaching to macroscopic level, where a quantized conductivity or resistivity of medium is noticed. These quantized aspects will  gradually disappear, when we go from high to low $B$, where the quantum identity of transportation is transformed to its classical expectation. From Fig.~(\ref{sig_QM}), one can notice that the fluctuations are crowded or congested more and more as we go to lower $B$. In other word, the interval of points in envelop reduces as we decrease the magnetic field, which means the changes of conductivity or resistivity are appeared to be continuous instead of quantized in nature. This transition from continuous to quantized values can be identified as classical to quantum ranges of $B$. The transition between the classical and quantum versions for increasing $B$ are all the more visible in the plots for $\sigma_{zz}^{\rm QM}$ and $\rho_{zz}^{\rm QM}$, as they have been scaled with $\sigma_0$ and $\rho_0$ respectively, i.e. the $B=0$ conductivity/resisitivity, which is again basically the $\sigma_{zz}$ and $\rho_{zz}$ for the classical case. 

	\begin{figure}  
		\centering	  
		\includegraphics[scale=0.35]{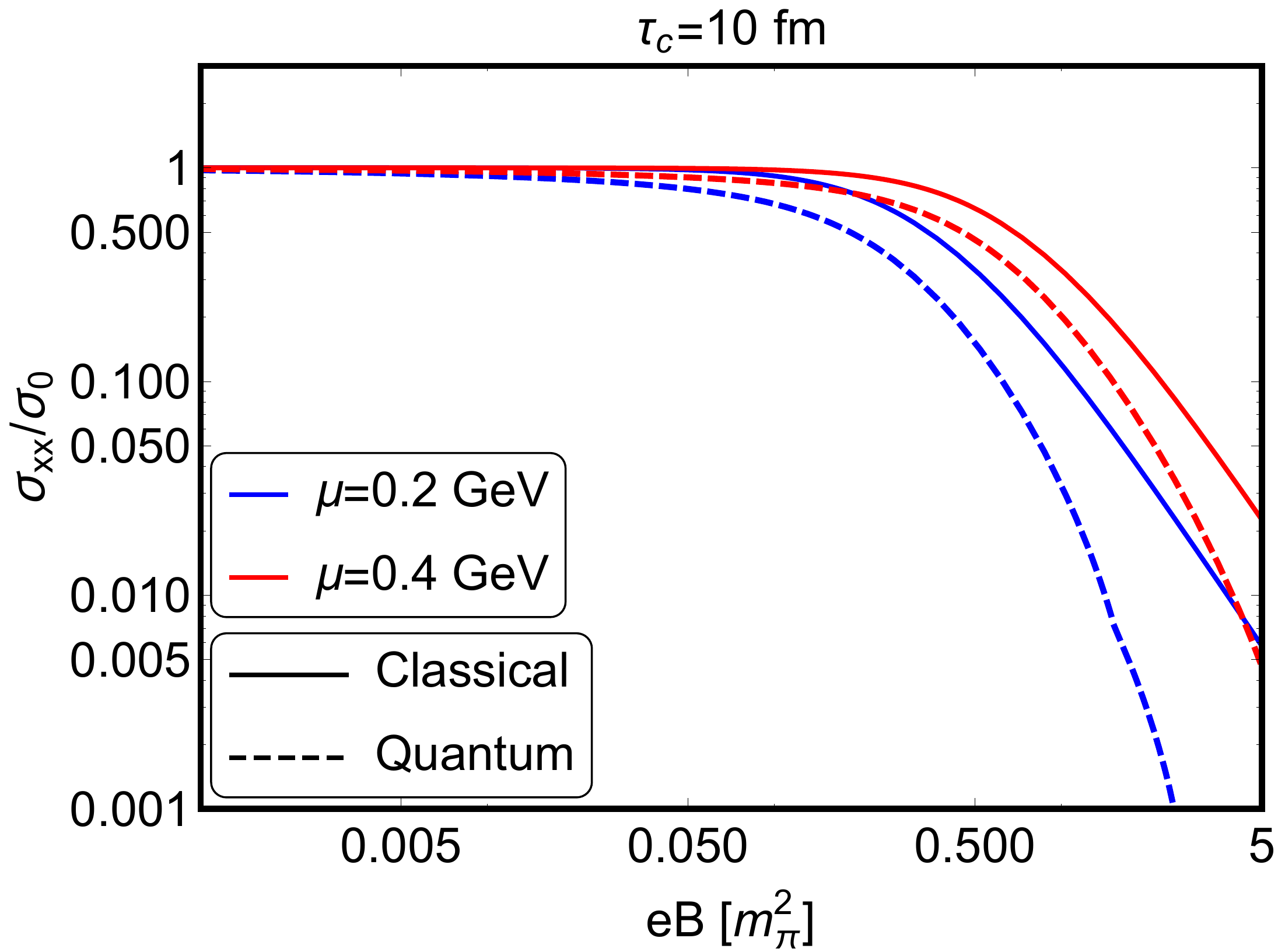} 
		\includegraphics[scale=0.32]{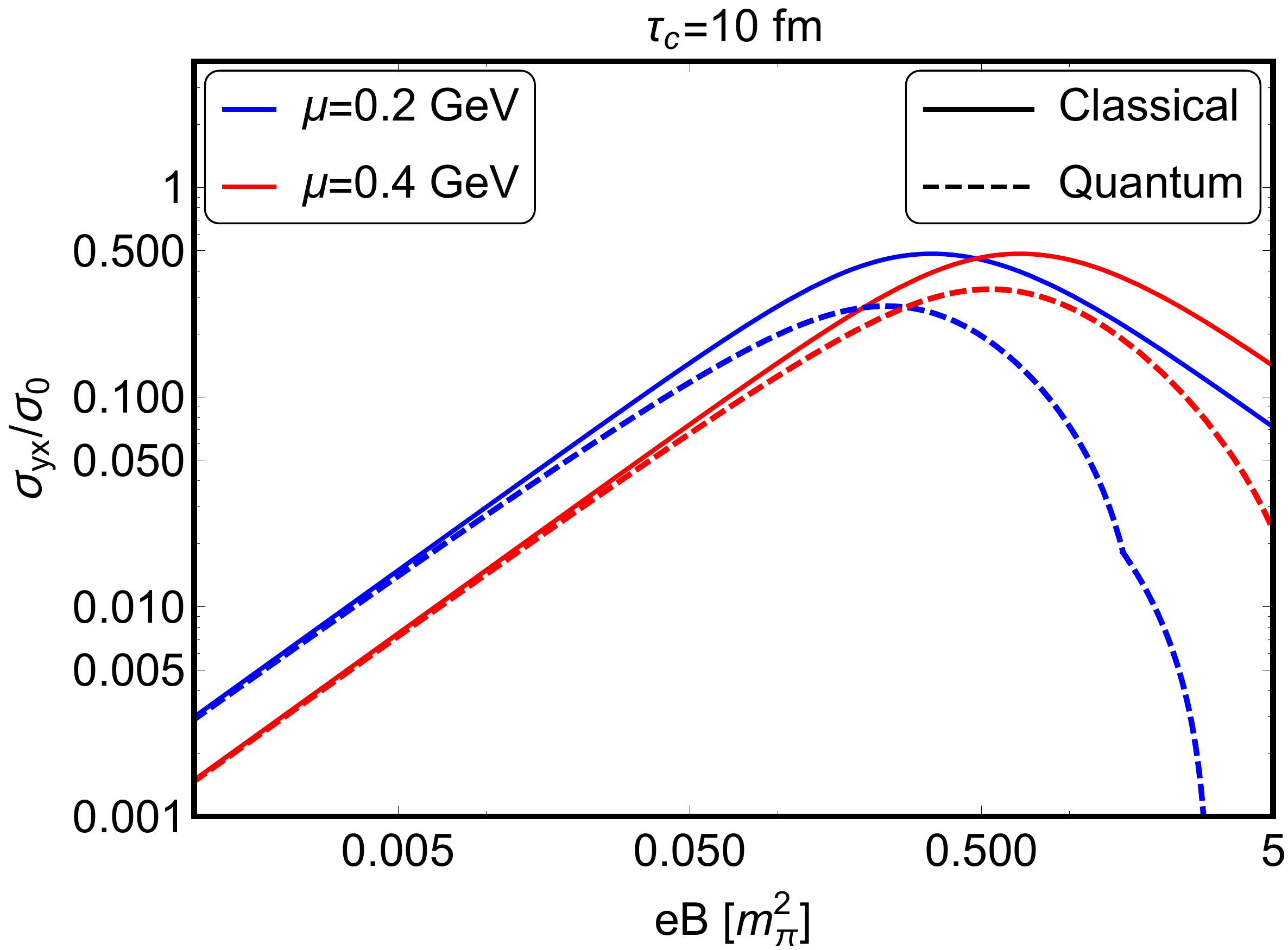}
		\includegraphics[scale=0.32]{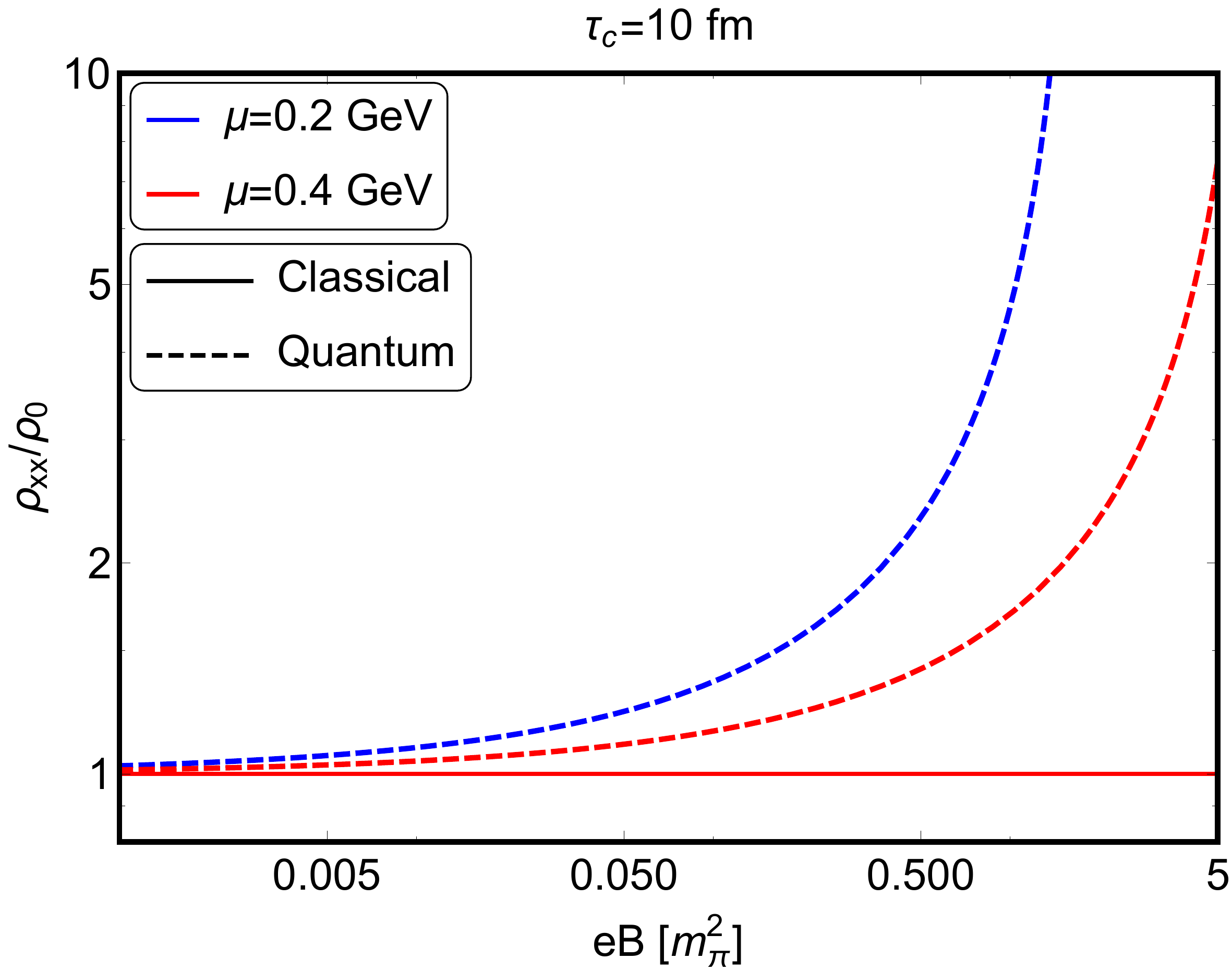}
		\includegraphics[scale=0.35]{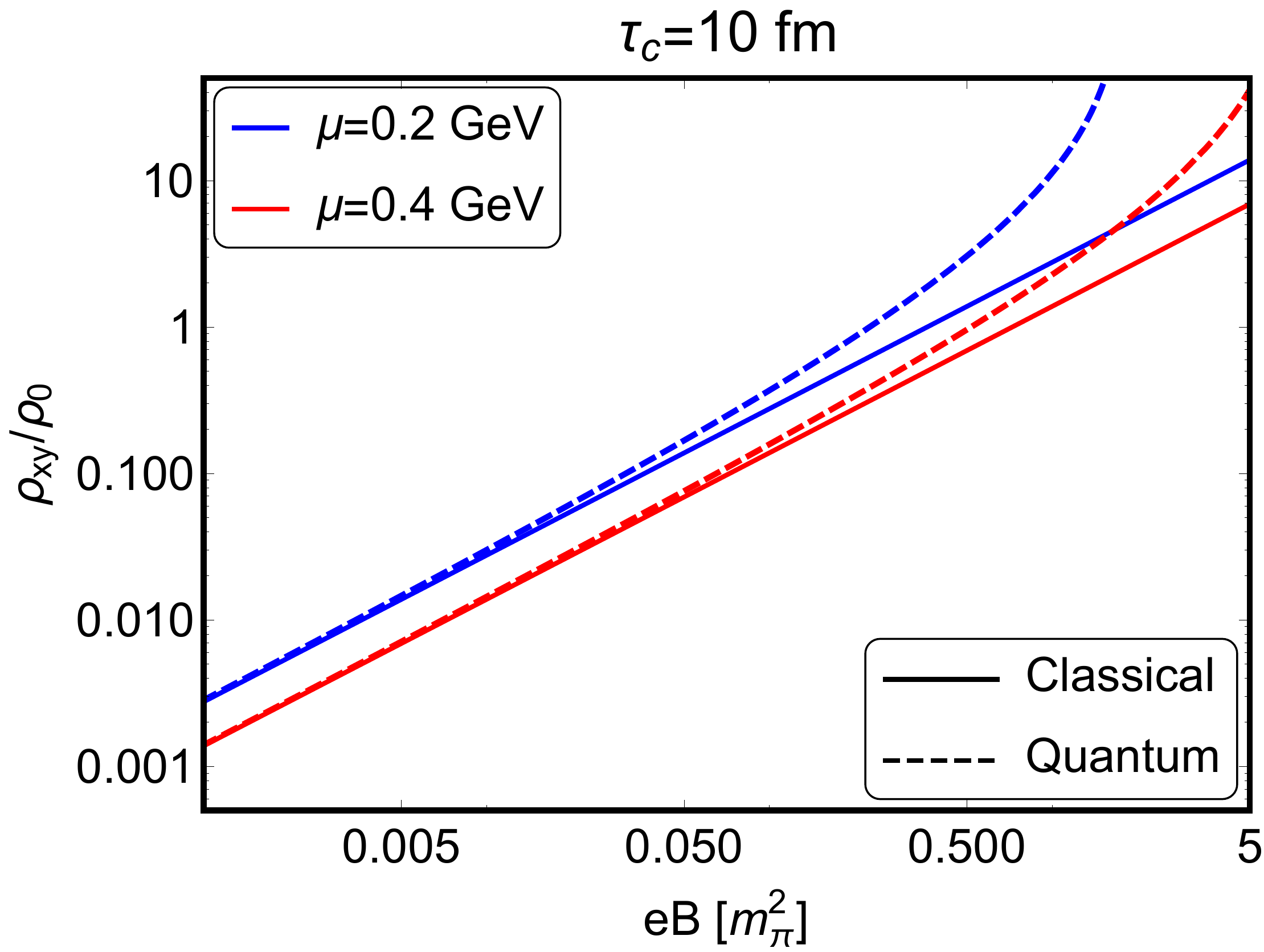}
		\caption{Classical (solid line) and quantum (dash line) curves of Perpendicular and Hall conductivity ($\sigma_{xx}$, $\sigma_{xy}$) and resistivity ($\rho_{xx}$, $\rho_{xy}$).  $\sigma_0$ and $\rho_0$ respectively signify the conductivity and resistivity for $eB=0$.  }
		\label{sig_QM2}
	\end{figure} 
	To visualize this classical to quantum transition in a more quantitative way, Fig.~(\ref{sig_QM2}) is dedicated to show the classical (solid line) and quantum (dash line) curves within the same portrait. Classical curves are based on Eqs.~(\ref{xy_F}) and (\ref{sF_cl}), while for quantum curves, Eq.~(\ref{sF_cl}) will be replaced only by Eqs.~(\ref{sD_QM_T0}) and \ref{sD_QM_T1}. Also instead of their exact numerical values, their envelop of $\zeta$-regularized values are presented.
	Analyzing the classical expressions (\ref{xy_F}), (\ref{sF_cl}), one can easily detect that the $B$-dependent anisotropic factors $\frac{1}{1+(\tau_c^Q/\tau_{BF})^2}$ and $\frac{(\tau_c^Q/\tau_{BF})}{1+(\tau_c^Q/\tau_{BF})^2}$ are basically responsible for monotonically decreasing trend and non-monotonically first increasing, then decreasing trend of $\sigma_{xx}$ and $\sigma_{xy}$ respectively. Envelop of quantum curves also follow more or less same pattern but they are deviated from each other. We might be focused on the deviation in large $B$ range, where quantum curves are quite suppressed than the classical curves and also distinguishable quantized values will be noticed if we take envelop points only. Based on Fig.~(\ref{lmax_B}), $l_{max}$ varies within $1-10$ in the ranges $e_QB\approx (0.1-1) m_\pi^2$ and $(1-5) m_\pi^2$ for $\mu=0.2$ GeV and $0.4$ GeV respectively, which can be roughly considered as their respective large $B$ ranges. Now, as we move towards smaller and smaller values of $e_QB$, we have to count larger and larger values of $l_{max}$. By those appropriate care of numerics, one can see that the quantum and classical curves will merge in a very low $B$ zone, which is shown in Fig.~\ref{sig_QM2} by using the zeta regularized values for the quantum curves. So, considering classical curves in low $B$ zone and quantum curves in high $B$ zone might be an alternative way to look at Fig.~(\ref{sig_QM2}), where deviation between them in high $B$ zone is our main focal interest.          
	
	On the other hand, the resistivity components of classical curves follow $\rho_{xx}(e_QB)=$ constant and $\rho_{xy}(e_QB)\propto e_QB$ trends, as noticed in the lower panels of Fig.~(\ref{sig_QM2}). Both are enhanced for quantum case in high $B$ zone. From classical curves, we can expect $\sigma_{xx, xy}\rightarrow 0$ and $\rho_{xx}=$constant, $\rho_{xy}\rightarrow \infty$ at $B\rightarrow \infty$, but for quantum curves, when we consider exact envelop points, we will get jumping-decreasing values of $\sigma_{xx, xy}$ and jumping-increasing values of $\rho_{xx, xy}$ upto the $B_0$. Beyond the value of $B_0$, $\sigma_{xx, xy}$ will suddenly drop to zero and in other hand, $\rho_{xx, xy}$ will suddenly blows up to infinity instead of any continuous dropping or blowing, expected in classical case. In this extreme scenario, fluid dynamics in perpendicular plane completely stop and only parallel direction to magnetic field has conduction with a saturated value of $\sigma_{zz}$, given in Eq.~(\ref{szz_l0}) or Eq.~(\ref{szz_lm0}).

	\section{Summary}
	\label{sec4}
	In summary, the conduction picture in presence of magnetic field
	is as follows. At zero temperature ($T=0$) and finite chemical potential $\mu$, conductivity components along x, y, and z are same $\sigma_{xx}(\mu)=\sigma_{yy}(\mu)=\sigma_{zz}(\mu)$ for without magnetic field case ($B=0$). At finite $B$, due to classically understood cyclotron motion, driven by Lorentz force, isotropic expectation of conduction is broken as $\sigma_{zz}(\mu)>\sigma_{xx}(\mu, B)=\sigma_{yy}(\mu, B)$, and also a Hall component $\sigma_{xy}(\mu, B)=-\sigma_{yx}(\mu, B)$ will be grown. According to classical expectation, at $B\rightarrow\infty$, maximum anisotropy will be built as $\sigma_{zz}(\mu)>\sigma_{xx}(\mu, B\rightarrow\infty)=\sigma_{yy}(\mu, B\rightarrow\infty)\rightarrow 0$ but in actual picture, Landau quantization will start to make impact on anisotropic conduction, when we go to the high $B$ zone. In that quantum picture, after a threshold magnetic field $B_0$, $\sigma_{xx}$ and $\sigma_{xy}$ can drop suddenly to zero and the extreme anisotropic picture (3D to 1D transformation) will be built.  
	The $\sigma_{zz}(\mu)$, which was independent of $B$ in classical case, will depend on $B$ due to Landau quantization. 
	Also, other components $\sigma_{xx}$, $\sigma_{xy}$ will be deviated from their classical curves in high $B$ zone. The decrements of conductivity components $\sigma_{xx}$, $\sigma_{xy}$ and increments of corresponding resistivity components $\rho_{xx}$, $\rho_{xy}$ with $B$ will follow the pattern of quantized jumps. This isotropic to anisotropic picture and classical to quantum pattern of conductivity and resistivity tensors for degerate massless quark matter is basically tried to sketched here by keeping in mind about quark core inside neutron star (NS). 
	
	Present work might be considered as a toy-level estimations for massless quark matter inside neutron star, which is pointing out a possibility of quantum Hall effect (QHE) pattern in NS. Though searching this QHE in NS within QCD time scale might be challenging task for experimental signal but there will be series of connecting issues, which might be very interesting future research plans by extending the present work. One of the agenda might be QHE pattern around the transition density or chemical potential, where quark-hadron phase transition is occured. Another immediate plan is to estimate other transport coefficients like shear viscosity, bulk viscosity, thermal conductivity etc. Also a rigorous quantum calculation of transport coefficients can be done via Kubo framework, which can say some additional fact in quantum field theoretical world as mentioned in Refs.~\cite{SG2,SS_SG2}.

	{\bf Acknowledgment:} Authors acknowledge to Aditya Rajesh for intial time involvement in numerical activity. JD, AG, NP and SG are benifitted from differen activities like journal clubs in IIT Bhilai, whose indirect impact on the present work should certainly be acknowledged.  A.B. acknowledges the support from Guangdong Major Project of Basic and Applied Basic Research No. 2020B0301030008 and Science and Technology Program of Guangzhou Project No. 2019050001.

	\appendix
	\section{Electrical conductivity for $B= 0$}
	\label{appA}
	Let us consider a relativistic charged gas, having a finite electrical conductivity. 
	By applying electric field ${\vec E}=E_x {\hat x}$ on the gas, we can measure the values
	of conductivity by measuring the current density ${\vec J}=J_x {\hat x}$,
	owing to the macroscopic Ohm's law
	\be
	J_x=\sigma_{xx}E_x~,
	\label{macro_E}
	\ee
	where $\sigma_{xx}$ is the electrical conductivity along x direction. In general,
	conductivity of electron gas should not depend on direction, which means that
	$\sigma_{xx}=\sigma_{yy}=\sigma_{zz}$ i.e. conductivity is isotropic in nature.
	We will see latter in this formalism part that magnetic field can destroy this
	isotropic nature, therefore, we start with the notation of direction indices (x, y, z),
	so that we can easily visualize a direction dependent conduction picture in presence of magnetic
	field.
	
	In microscopic ideal picture of electron gas, we assume that 
	they are in thermal equilibrium by following the distribution function,
	\be
	f_0=\frac{1}{e^{\beta(\om -\mu)}+ 1}~,
	\ee
	carrying information of medium like temperature $T=\beta^{-1}$, chemical potential $\mu$
	and information of medium constituents like energy of constituent particle $\om=\{\vk^2+m^2\}^{1/2}$.
	One can assume roughly a zero net current density
	$J_x =ge\int \frac{d^3k}{(2\pi)^3}\frac{k_x}{\om}f_0\approx 0$
	in this ideal picture. When we apply external electric field,
	we can assume a new distribution $f=f_0 +\delta f$, which is
	quite close to equilibrium distribution with a small deviation $\delta f$.
	Hence, we will get a non-zero net current density (due to external
	electric field)
	\be
	J_x =g e\int \frac{d^3k}{(2\pi)^3}\frac{k_x}{\om}\delta f~.
	\label{J_df}
	\ee
	The $\delta f$ is completely originated for external electrical field $E_x$
	and their link can be found by
	using relaxation time approximatio (RTA) of Boltzmann equation (BE)
	\bea
	\frac{\del f}{\del t}+\frac{\del x^i}{\del t}\frac{\del f}{\del x^i}+\frac{\del k^i}{\del t}\frac{\del f}{\del k^i}
	&=&-\frac{\delta f}{\tau_c}
	\nn\\
	\Rightarrow\frac{\del f_0}{\del t}+v^i\frac{\del f_0}{\del x^i}+F^i\frac{\del f_0}{\del k^i}
	&=&-\frac{\delta f}{\tau_c}~,
	\nn\\
	\label{BE}
	\eea
	where we have assumed $f\approx f_0$ in the left hand side (LHS) of Eq.~(\ref{BE}).
	Now, external electric field acts as a force term $F^i$, which is solely responsible
	for small deviation $\delta f$ within a time $\tau_c$ (known as relaxation time). So
	the first two terms of LHS of BE will not contribute for non-zero $\delta f$,
	only third term will be responsible and thus Eq.~(\ref{BE}) becomes
	\bea
	e{\vec E}\cdotp{\vec \nabla_k} f_0&=&-\delta f/\tau_c
	\nn\\
	\Rightarrow \delta f &=& -\tau_c{e}{\vec E}\cdotp\frac{\vec k}{\om}
	\Big[\frac{\del f_0}{\del \om}\Big]
	\nn\\
	&=&\tau_c{e}E_x\Big(\frac{k_x}{\om} \Big)[\beta f_0(1- f_0)]~,
	\label{del_f}
	\eea
	Using Eq.~(\ref{del_f}) in Eq.~(\ref{J_df}) and then comparing with Eq.~(\ref{macro_E}),
	we will get
	\bea
	J_x &=&\Big[g{e^2}\beta\int \frac{d^3k}{(2\pi)^3}\frac{k^2_x}{\om^2}\tau_c f_0(1- f_0)\Big ]E_x
	\nn\\
	\Rightarrow \sigma_{xx} &=& g{ e}^2 \beta\int \frac{d^3k}{(2\pi)^3}
	\tau_c\frac{k_x^2}{\om^2} f_0(1- f_0)~.
	\label{micro_E}
	\eea
	Similar to $x$-direction, if we repeat the calculation for $y$ and $z$-direction, 
	then we can get the isotropic expressions 
	\bea
	\sigma_{xx}=\sigma_{yy}=\sigma_{zz}=g{e}^2 \beta\int \frac{d^3k}{(2\pi)^3}
	\tau_c\frac{\vk^2}{3\om^2} f_0(1- f_0)~.
	\eea
	
	\section{Electrical conductivity for $B\neq 0$}
	\label{appB}
	
	Let us assume an external
	magnetic field ${\vec B}=B {\hat z}$ is applied on the relativistic electron gas.
	Hence, the force term $\frac{d\vk}{dt}={ e} ({\vec E} +{\vec v}\times {\vec B})$ in 
	Eq.~(\ref{BE}) will be replaced as
	\bea
	{e} ({\vec E} +\frac{\vec k}{\om}\times {\vec B})\cdot\nabla_k f_0 &=& \frac{-\delta f}{\tau_c}
	\nn\\
	{e} ({\vec E} +\frac{\vec k}{\om}\times {\vec B})\cdot\Big(\frac{\vk}{\om}\Big)\frac{\del f_0}{\del\om} 
	&=& \frac{-\delta f}{\tau_c}~.
	\eea
	The second term of the LHS is $B$ dependent term and will be vanished
	due to the vector identity $(\vk\times{\vec B})\cdot \vk={\vec B}\cdotp(\vk\times\vk)=0$.
	So to include a $B$ dependent term, we consider the $\nabla_k(\delta f)$ term also
	in BE,
	\be
	{e}{\vec E}\cdot\Big(\frac{\vec k}{\om}\Big)\frac{\partial f_0}{\partial \om} 
	+ {e}(\frac{\vec k}{\om}\times {\vec B})\cdot \nabla_k(\delta f)
	=-\delta f/\tau_c~,
	\label{RBE_H_df}
	\ee
	where without field ansatz, given in Eq.~(\ref{del_f}), will now be considered as
	\bea
	\delta f &=& -\phi\frac{\del f_0}{\del \om}
	\nn\\
	&=& \Big(\vk\cdot {\vec F}\Big)\beta f_0(1-f_0)~,
	\label{df_B}
	\eea
	with 
	\be
	{\vec F}= (A_x {\hat x} + A_z{\hat z} + A_y({\hat x}\times{\hat z}) )~.
	\label{F_A}
	\ee
	So reader can understand that for $B=0$ case, we have ${\vec F}= A_x {\hat x}=\alpha E_x {\hat x}$
	but for $B\neq 0$ case, we have two more components along the direction of ${\vec B}=B {\hat z}$,
	and ${\vec E}\times{\vec B}=E_x B {\hat x}\times{\hat z}$. Our goal is to find the coefficients
	$A_x$, $A_y$ and $A_z$, associated with the three directions. 
	
	Now, using Eqs.~(\ref{df_B}), (\ref{F_A}) in Eq.~(\ref{RBE_H_df}), we get
	\bea
	\Big(\frac{\vec k}{\om}\Big)\cdot\Big[{e}{\vec E} - {e}({\vec B}\times{\vec F})\Big]
	&=&\vk\cdot {\vec F}/\tau_c
	\nn\\
	\Rightarrow \frac{\tau_c}{\om}\Big[{e}E_x{\hat x} - {e}B{\hat z}\times\Big(A_x {\hat x} + A_z{\hat z} 
	- A_y{\hat y}\Big)\Big]
	&=&\Big(A_x {\hat x} + A_z {\hat z} - A_y {\hat y}\Big)
	\label{RTA_RBE_B}
	\eea
	where we have used a standard vector identity,
	\bea
	(\frac{\vec k}{\om}\times {\vec B})\cdot\nabla_k(\delta f)
	&=&-(\frac{\vec k}{\om}\times {\vec B})\cdot \nabla_k(\vk\cdot {\vec F}) \frac{\partial f_0}{\partial \om}
	\nn\\
	&=&-\frac{\vec k}{\om}\cdot ({\vec B}\times{\vec F}) \frac{\partial f_0}{\partial \om}~.
	\label{del_f_expand}
	\eea
	The coefficients of ${\hat x}$, ${\hat z}$ and ${\hat y}$ in Eq.~(\ref{RTA_RBE_B}) 
	will give us relations
	\bea
	A_z &=& 0
	\nn\\
	A_x &=& \frac{1}{1+(\tau_c/\tau_B)^2}\frac{\tau_c}{\om}E_x
	\nn\\
	A_y &=& \frac{-\tau_c/\tau_B}{1+(\tau_c/\tau_B)^2}\frac{\tau_c}{\om}E_x~,
	\eea
	where $\tau_B=\om/(eB)$ is inverse of synchrotron frequency.
	Hence, Eq.~(\ref{df_B}) can be written as
	\be
	\delta f ={e}\tau_c\Big(\frac{k_x}{\om} - \frac{k_y}{\om}\frac{\tau_c}{\tau_B}\Big)E_x
	\frac{1}{1+(\tau_c/\tau_B)^2}~\beta f_0(1-f_0)~,
	\ee
	which guide us to think about a matrix form of Ohm's law,
	\be
	\left(
	\begin{array}{c}
		J_x \\
		J_y 
	\end{array}
	\right) =
	\left(
	\begin{array}{cc}
		\sigma_{xx} & \sigma_{xy} \\
		\sigma_{yx} & \sigma_{yy}
	\end{array}
	\right)
	\left(
	\begin{array}{c}
		E_x \\
		0 
	\end{array}
	\right)~,
	\label{cond_matrix}
	\ee
	where
	\bea
	\sigma_{xx}&=&g{e}^2 \beta\int \frac{d^3k}{(2\pi)^3}
	\tau_c\frac{1}{1+(\tau_c/\tau_B)^2}\frac{k_x^2}{\om^2} f_0(1- f_0)
	\\
	\sigma_{yx}&=&-g{e}^2 \beta\int \frac{d^3k}{(2\pi)^3}
	\tau_c\frac{\tau_c/\tau_B}{1+(\tau_c/\tau_B)^2}\frac{k_y^2}{\om^2} f_0(1- f_0)~.
	\nn\\
	\eea
	By applying ${\vec E}=E_y{\hat y}$, a similar calculation will give us
	the expressions of $\sigma_{yy}$, $\sigma_{xy}$, which follow the relations 
	$\sigma_{xx}=\sigma_{yy}$, $\sigma_{xy}=-\sigma_{yx}$. 
	Longitudinal conductivity along z-axis will remain unaffected by magnetic field as
	\be
	\sigma_{zz}=g{ e}^2 \beta\int \frac{d^3k}{(2\pi)^3}
	\tau_c\frac{k_z^2}{\om^2} f_0(1- f_0)~.
	\ee
	Without magnetic field picture resistivity $\rho_D$ will be exactly inverse of conductivity $\sigma$
	i.e. $\rho=1/\sigma$ but in presence of magnetic field, additional Hall conductivity ($\sigma_{xy}=-\sigma_{yx}$) 
	will build a conductivity matrix, as defined in Eq.~(\ref{cond_matrix}). Inverse matrix of that will give
	resistivity matrix~\cite{Tong} with inverse relation of Eq.~(\ref{cond_matrix}):
	\be
	\left(
	\begin{array}{cc}
		\rho_{xx} & \rho_{xy} \\
		\rho_{yx} & \rho_{yy}
	\end{array}
	\right)
	\left(
	\begin{array}{c}
		J_x \\
		J_y 
	\end{array}
	\right) =
	\left(
	\begin{array}{c}
		E_x \\
		0 
	\end{array}
	\right)~.
	\label{rho_matrix}
	\ee
	Comparing Eq.~(\ref{cond_matrix}) and (\ref{rho_matrix}), we get the relations for resistivity as
\bea
\rho_{xx} &=& \rho_{yy} = \frac{\sigma_{xx}}{\sigma_{xx}^2
	+ \sigma_{xy}^2},
\nn\\
\rho_{xy} &=& - \rho_{yx} = -\frac{\sigma_{xy}}{\sigma_{xx}^2
	+ \sigma_{xy}^2} = \frac{\sigma_{yx}}{\sigma_{xx}^2
	+ \sigma_{xy}^2}~,
\nn\\
 \rho_{zz} &=& \frac{1}{\sigma_{zz}}~.
\label{rho-sigma}
\eea

\end{document}